\documentclass[runningheads]{llncs}
\usepackage[T1]{fontenc}
\usepackage{graphicx}
\usepackage{multirow}
\usepackage{cite}
\usepackage{amsmath,amssymb,amsfonts}
\usepackage{algorithmic}
\usepackage{graphicx}
\usepackage{textcomp}
\usepackage{xcolor}
\usepackage[ruled,linesnumbered]{algorithm2e}
\usepackage{subfigure}
\usepackage{booktabs}
\usepackage{multirow}
\usepackage{array}
\usepackage{arydshln}
\usepackage{enumitem}
\usepackage{setspace}
\usepackage{listings}
\begin{document}
%
\title{Symbolic \& Acoustic: Multi-domain Music Emotion Modeling for Instrumental Music}
%
\titlerunning{Multi-domain Music Emotion Modeling for Instrumental Music}

\author{Kexin Zhu$^\ddagger$, Xulong Zhang$^\ddagger$, Jianzong Wang\thanks{Corresponding author: Jianzong Wang, jzwang@188.com. $^\ddagger$ These authors are equal contribution.}, Ning Cheng, Jing Xiao }
\authorrunning{K. Zhu et al.}

\institute{Ping An Technology (Shenzhen) Co., Ltd.}
\maketitle              
\begin{abstract}
  Music Emotion Recognition involves the automatic identification of emotional elements within music tracks, and it has garnered significant attention due to its broad applicability in the field of Music Information Retrieval. It can also be used as the upstream task of many other human-related tasks such as emotional music generation and music recommendation. Due to existing psychology research, music emotion is determined by multiple factors such as the Timbre, Velocity, and Structure of the music. Incorporating multiple factors in MER helps achieve more interpretable and finer-grained methods. However, most prior works were uni-domain and showed weak consistency between arousal modeling performance and valence modeling performance. Based on this background, we designed a multi-domain emotion modeling method for instrumental music that combines symbolic analysis and acoustic analysis. At the same time, because of the rarity of music data and the difficulty of labeling, our multi-domain approach can make full use of limited data. Our approach was implemented and assessed using the publicly available piano dataset EMOPIA, resulting in a notable improvement over our baseline model with a 2.4\% increase in overall accuracy, establishing its state-of-the-art performance.

\end{abstract}
\section{Introduction}

The emotional aspect of music, commonly known as its affective content, holds significant importance and is often regarded as the essence of musical expression. The recognition of emotions in music, known as Music Emotion Recognition (MER), has emerged as a prominent topic and crucial objective within the field of Music Information Retrieval (MIR). This recognition process assumes paramount significance due to its widespread application in various scenarios involving emotion-driven music retrieval and recommendation. Restricted by the complexity of emotion, research on MER has encountered great difficulties\cite{zongshu,ru2023Improving}. Emotion is a very complex psychological state, and different people have different emotional thresholds~\cite{tang2023EmoMix}. This makes emotional annotation more difficult and emotional data more scarce.

The recognition and understanding of the intricate interplay between various factors within music and their impact on music emotion constitute a central concern in ongoing research on MER. Investigating this matter not only facilitates the advancement of more efficient and nuanced MER techniques but also contributes to the development of comprehensive insights into the complex nature of music emotion. Existing research usually applies disentanglement or multi-domain analysis to modeling music emotion from multiple aspects. Berardinis et al. \cite{multiple} applies Music Source Separation during pre-processing and analyze the emotional content in vocal, bass, drums, and other parts separately, their proposed method shows promising performance. Zhao et al. \cite{icme} provide a new perspective by modeling music emotion with both music content and music context, their proposed method applies multi-modal analysis on audio content, lyrics, track name, and artist name of the music. 
								
To further explore the essence of music emotion, research was also carried out on instrumental music. In the field of psychology and affective computing, Laukka et al. \cite{laukka} proposed a convincing music emotion perception model for instrumental music and concluded six factors that affect music emotion: Dynamics, Rhythm, Timbre, Register, Tonality, and Structure.  Those factors reflect both the acoustic characteristics and the structural characteristics of the music.  Laukka's model indicates the importance of incorporating both acoustic analysis and symbolic analysis for MER. Acoustic factors such as Dynamics and Timbre are highly related to the Arousal expression of the music but are not included in the symbolic representations of music. Therefore symbolic-only methods show relatively weaker performance on Arousal detection. 
Structural factors such as Tonality and Structure are highly related to the Valence expression. Although those factors are included in the acoustic domain, existing acoustic analysis methods can hardly learn the structural information without extra supervision. To incorporate all the important factors, both acoustic analysis and symbolic analysis are needed.

However, most existing MER methods for instrumental music are uni-domain and fail to model music emotion from multiple aspects.  Existing researches mainly apply deep-learning-based methods on the acoustic domain or uses sequence-modeling methods on the symbolic domain representations of the music. In their recent publication on emotion recognition in symbolic music, Qiu et al. \cite{symbolic} introduced a pioneering approach utilizing the MIDIBERT model \cite{midibert}, a large-scale pre-trained music understanding model. At present, no existing research on Music Emotion Recognition (MER) for instrumental music integrates both acoustic and symbolic analyses. As a result, we present an innovative method in this study that encompasses music emotion modeling from both acoustic and symbolic perspectives. Given the representative nature of piano music within the instrumental domain, we implemented and conducted an evaluation of our proposed approach using the publicly available piano emotion dataset EMOPIA \cite{emopia}.

Our contribution can be summarized as follows:
\vspace{-4pt}
\begin{itemize}
\item Inspired by existing psychology and affective computing research, we proposed a multi-domain emotion modeling method for instrumental music, which only needs audio input.
Our method used a pre-trained transcription model to obtain symbolic representation, therefore can be used on each instrument that can be automatically transcripted.

\item We designed a refined acoustic model with mixed acoustic features input and a transformer-based symbolic model. Both models showed promising performance.

\item We implemented and evaluated our proposed method on the public piano emotion dataset EMOPIA \cite{emopia}. Our method achieved state-of-the-art performance on EMOPIA with better consistency between Valence detection and Arousal detection performance.

\end{itemize}

\section{Related Works}
\label{sec:related}
There have been many studies in the research field of MER. According to the different domains of focus, these studies include MER with acoustic-only and MER with symbolic-only studies. These works have promoted progress in MER, and there are also some points that can be improved.

\subsection{MER with Acoustic-only}
In order to explore which part of the vocal or accompaniment music carries more emotional information, Xu et al. \cite {xu2014source} used the sound source separation technology, combined with the 84-dimensional manual low-level features (such as Mel frequency cepstrum coefficient (MFCC), spectral center, spectral attenuation point, spectral flux, and other similar measures.), and then used a classifier to recognize music emotion. 
Coutinho et al. \cite{coutinho2015automatically} extracted 65 Low-level Descriptors (LLDs) in a time window of 1 second and calculated their first-order difference to obtain a total of 130 low-level features, then calculated the mean and standard deviation of each LLD in one second, and finally formed a 260-dimensional feature vector, and then used Long Short-term Memory (LSTM) network to carry out regression prediction of dynamic V/A (Valence / Arousal) value. 
Fukayama et al. \cite{fukayama2016music} proposed a method to adapt to aggregation by considering new acoustic signal input based on multi-stage regression. At the same time, a method of adjusting the aggregation weight is introduced to deal with the emotion caused by the new input that cannot be known in advance, and the deviation observed in the training data is utilized by using Gaussian process regression. 
Li et al. \cite{li2016dblstm} introduced a novel approach to tackle dynamic emotion regression by leveraging Deep Bi-directional Long Short-term Memory (DBiLSTM) in a multi-scale regression framework . 
Moreover, the author also examined the influence of dissimilar sequence lengths between the training and prediction stages on the overall performance of DBiLSTM. By investigating this aspect, the study aimed to gain insights into the effects of such variations on the efficacy of the model.
\cite{malik2017stacked} uses the CNN network that can process local information with fewer parameters and the RNN network that can process context information, that is, the CRNN structure, which uses the least parameters than Media Eval 2015. 

Other methods have achieved the best results in the dynamic regression prediction of emotion at that time. Huang et al. \cite{huang2017music} introduced the attention mechanism into the music emotion classification task, and introduced the attention layer with short-term and short-term memory units into the deep convolution neural network for music emotion classification. Different weights are allocated on different time blocks (chunks), and the song-level emotion classification prediction is obtained through fusion. Liu et al. \cite{liu2017cnn} regards music emotion recognition as a multi-label classification task, and uses convolutional neural networks and spectrum diagram to complete end-to-end classification. Chen et al. \cite{chen2017high} considered the complementarity between CNN with different structures and between CNN and LSTM, and combined multi-channel CNN with different structures and LSTM into a unified structure (Multi-channel Convolutional LSTM, MCCLSTM) to extract advanced music descriptors.
Choi et al. \cite{choi2017transfer} employed a pre-trained convolutional neural network (CNN) feature, which was initially trained for music auto-tagging purposes. They then successfully transferred this CNN to various music-related classification and regression tasks, showcasing its adaptability and versatility. Similarly, Panda et al. \cite{panda2018musical} introduced a collection of innovative affective audio features to enhance emotional classification in audio music. The authors observed that conventional feature extractors primarily focus on low-level timbre-related aspects, neglecting essential elements like musical form, texture, and expressive skills. To address this limitation, the authors devised a novel set of algorithms specifically designed to capture information related to music texture and expression, effectively compensating for the significant gaps in music emotion recognition research.

\subsection{MER with Symbolic-only}
Previous research employed manual extraction of statistical musical characteristics, which were subsequently inputted into machine learning classifiers to forecast the emotional aspects of notated music. Grekow et al. \cite{grekow2009detecting} conducted an analysis on classical music in MIDI format and extracted 63 distinct features. In a similar vein, Lin et al. \cite{lin2013exploration} conducted a comparative investigation involving multiple features (audio, lyrics, and MIDI) extracted from the same music. Remarkably, they discovered that MIDI features exhibited superior performance in emotion recognition. Building upon this finding, the researchers utilized the JSymbolic library \cite{mckay2006jsymbolic} to extract 112 advanced music features from MIDI files. Subsequently, Support Vector Machine (SVM) was employed to classify the data. Similarly, Panda et al. \cite{panda2013multi} employed various tools to extract features from MIDI files and utilized SVM for classification purposes.

More recent studies demonstrate a growing adoption of a symbolic music encoding technique similar to MIDI \cite{oore2020time}, which is gaining popularity among researchers. Additionally, deep learning models have emerged as the prominent approach in this field. Ferreira \cite{ferreira2021learning} devised a method to encode MIDI files into MIDI-like sequences, leveraging LSTM and GPT2 for sentiment classification purposes. This approach offers simplicity and efficiency. Drawing inspiration from the remarkable achievements of BERT, Chou et al. \cite{chou2021midibert} introduced MidiBERTPiano, a large-scale pre-trained model utilizing CP representation. The proposed model showcases promising outcomes in various domains, including symbolic music emotion recognition. Highlighting the paramount importance of emotional expression in music's intrinsic structure, Liu et al. \cite{qiu2022novel} proposed a straightforward multi-task framework for the symbolic MER task. Notably, this approach benefits from readily available labels for auxiliary tasks, eliminating the need for manual annotation of labels beyond emotion classification.

\section{Methodology}
\begin{figure*}[h]
  \centering
\includegraphics[width=\textwidth]{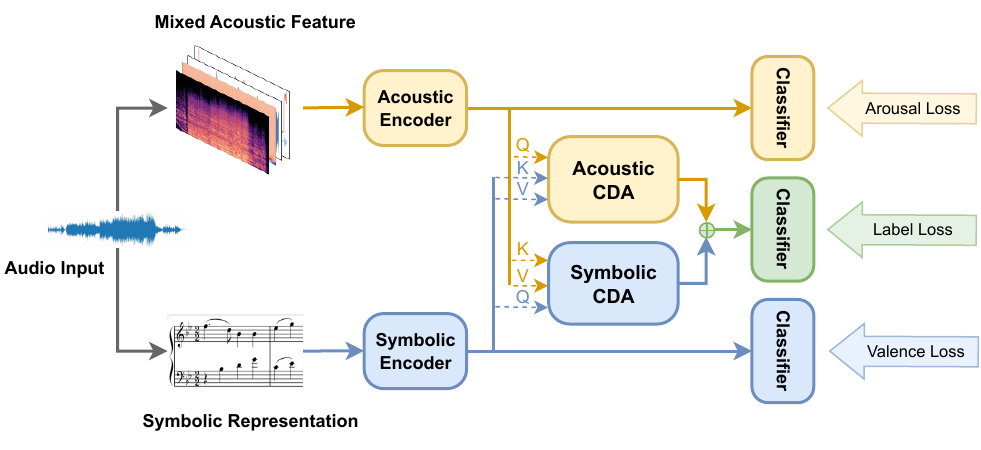}
  \caption{The overall structure. The feature representations of the two domains are generated by the acoustic domain branch and the symbolic domain branch in the model, and the fusion process is completed in CDA.}
  \label{model2}
\end{figure*}

The complete diagram illustrating the overall architecture of our proposed approach can be observed in Figure \ref{model2}. The structure contains two branches: the acoustic domain branch (marked in yellow) applies acoustic analysis on mixed acoustic features with a Conv-based acoustic encoder, and the symbolic domain branch (marked in blue) applies symbolic analysis on music score sequence by using a Transformer-based symbolic encoder. It is worth noting that the outputs of the two branches come from the same modality, that is, from the acoustic input, so they belong to different domains of the same modality.

\subsection{Acoustic Domain Analysis for Arousal Modeling}

For the acoustic domain analysis, we want to explicitly extract the information that relates to music emotion expressions, such as Timbre and Dynamics \cite{laukka}. We use a mixed feature as input, which consists of the Mel-frequency Cepstral Coefficient (MFCC), Mel-spectrogram, Spectral Centroid (SC), and Root Mean Square Energy (RMSE) of the audio input. SC and RMSE reflect the energy distribution and changes of the audio, which is strongly correlated to music emotion expression. We use mel-spectrogram instead of STFT spectrogram because it better fits the human auditory perception process. 
We also calculate a 20-dimensional MFCC with librosa\cite{librosa}.
After these features are obtained, we resize and align them in the time dimension. The mixed feature can be obtained by splicing these features.  

The processing flow of the acoustic domain branch is shown at the top of Figure \ref{model2}. We use a 2D-ConvNet module as the acoustic encoder for its great ability to encode temporal and frequency domain information simultaneously. 
After the feature extraction process, the extracted features are flattened and combined in the channel dimension to form the acoustic domain output. A comprehensive summary of the settings used in the experiment can be found in Table \ref{setting}.

Acoustic domain analysis shows better performance on Arousal detection than symbolic domain analysis. Arousal is mainly decided by acoustic attributes such as Dynamics, Energy, and Timbre, which are not included in symbolic domain representation. Therefore we calculate an extra arousal classification loss function using Binary Cross Entropy (BCE) on the acoustic domain analysis branch during the training process.

\subsection{Symbolic Domain Analysis for Valence Modeling}
As mentioned above, our proposed method is designed to perform both acoustic and symbolic domain analysis with only audio input. That is to say, our symbolic part uses the automatic piano transcription module to form the symbolic domain representation instead of directly using the MIDI files in the EMOPIA dataset. This provides a common paradigm for other transcribable musical instruments. Therefore for the symbolic domain analysis branch, we use a pre-trained automatic transcription model to perform piano transcription. Specifically, we use the refined version of Onsets and Frames \cite{onsets,onsets2} proposed by Zhao et al. \cite{csmt}, which shows better generalizability and costs fewer computation resources. The transcripted piano score is converted into MIDI format, which includes the onset, offset, duration, and velocity of each note. 

The music score is the "language" of the music and is a semantic sequence similar to natural language. Therefore the symbolic representation of the music score is similar to that of the natural language.

\begin{figure}[h]
  \centering
  \includegraphics[width=0.97\textwidth]{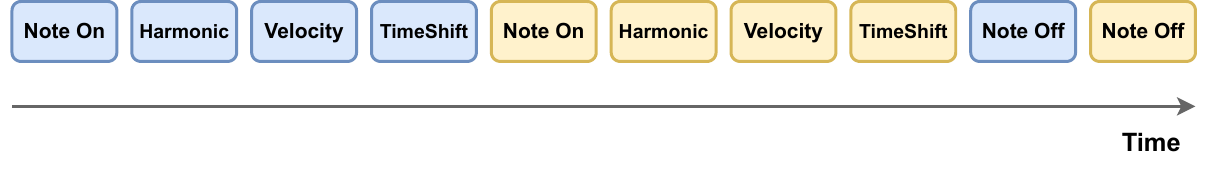}
  \caption{The refined MIDI-like symbolic representation we used.}
  \label{midi}
\end{figure}

In this work, we use a refined MIDI-like representation for note embedding, which is shown in Figure \ref{midi}. Unlike the original MIDI-like \cite{emopia39} representation, we add an attribute named "harmonic" which explicitly denotes the number of sounding notes at the onset of a note. Since harmonic is an important part of musical performance, we decide to add extra information about it. Therefore, the symbolic domain representation for a single note consists of the onset time, harmonic, velocity, time shift, and offset time of the note.

The structure of the symbolic domain analysis branch is shown at the bottom of Figure \ref{model2}. After the note embeddings are obtained, we input them into a Transformer encoder module\cite{transformer} to extract the emotional representation of the piano score. The Transformer encoder module consists of four original Transformer encoder layers adopted in \cite{transformer}. 
We pre-trained the encoder with the MIDI data from the MAESTRO dataset, for there are not enough samples in EMOPIA to train our Transformer encoder module.

Symbolic domain analysis mainly focuses on the high-level semantics of the note sequences, which leads to better Valence detection accuracy than acoustic analysis. As we want to make use of its advantage, we calculate an extra valence classification loss on the symbolic domain analysis branch during the training process.

\subsection{Combining Symbolic and Acoustic Analysis}
The final purpose of our method is to perform 4-Quadrant (4Q) classification concerning both Arousal and Valence, therefore the cross-domain feature fusion method is important.
When combining extracted acoustic domain features and symbolic domain features, the Cross-domain Attention (CDA) module is used for cross-domain feature fusion. CDA has a similar mechanism to multi-head cross-modal attention\cite{cma}. In CDA module, Query and Key-Value pairs come from two different domains instead of different modalities in cross-modal attention. Each attention head can be calculated separately:
\begin{eqnarray}
Attention(F_{Q},F_{K},F_{V})=softmax(\frac{F_{Q} (F_{K})^{T}}{\sqrt{d}})F_{V}\nonumber \\
~=softmax(\frac{F_{\alpha}W_{Q}(F_{\beta}W_{K})^{T}}{\sqrt{d}})F_{\beta}W_{V}
\label{atten}
\end{eqnarray}

Let $F_{Q}$, $F_{K}$, and $F_{V}$ denote the vectors for $Query$, $Key$, and $Value$, respectively. Within the attention mechanism, these input vectors are obtained by multiplying the extracted features of the $\alpha$ and $\beta$ domains, represented as $F_{\alpha}$ and $F_{\beta}$, with their respective learnable weight matrices $W_{Q}$, $W_{K}$, and $W_{V}$. Here, $d$ represents the dimension size of the $Key$ vector. The multi-head attention can be defined as the concatenation of each individual head:
\begin{eqnarray}
MultiHead(F_{\alpha},F_{\beta})= Concat(head_{1}, ..., head_{H})W_{O}\\
head_{i}=Attention(F_{\alpha}W_{Q}^{i},F_{\beta}W_{K}^{i},F_{\beta}W_{V}^{i})
\label{mha}
\end{eqnarray}

The learnable weight matrix $W_{O}$ and the number of attention heads $H$ play crucial roles in this multi-head attention mechanism. By leveraging multiple attention heads, this mechanism effectively highlights the significant aspects of each domain, which cannot be achieved through simple concatenation alone.

As shown in Figure \ref{model2}, in each processing procedure, our model calculates the CDA mechanism twice. We calculate an acoustic cross-domain attention mechanism and a symbolic cross-domain attention mechanism separately. This bidirectional CDA fusion strategy brings higher fusing efficiency. The output of acoustic CDA and symbolic CDA are concatenated and input into a classifier for 4Q emotion classification. During the training process, we calculate a 4Q Label loss on this classifier using Cross Entropy (CE) loss function.

\section{Experiments}
To assess the effectiveness of our proposed model, we conducted two primary types of experiments in this study: comparative studies and ablation studies. These experiments were designed to thoroughly evaluate and analyze the performance of our model from different perspectives.

\subsection{Expriments Setup}
We use the EMOPIA \cite{emopia} dataset, which is an open-source dataset for piano-based emotion recognition.
EMOPIA contains 1087 piano clips from 387 songs, all piano clips are annotated with their MIDI files and emotion labels. As only music metadata is available, we collect all music files by their corresponding YouTube ID with the 'youtube-dl' package. Following the configuration employed in \cite{emopia}, the dataset was divided into train-validation-test splits with a ratio of 7:2:1, ensuring appropriate proportions for training, validation, and testing stages. However, due to the unavailability of several music pieces on YouTube, we're only able to use approximately 90\% data of the whole dataset. Similarly, we not only perform the classification of 4 quadrants but also carry out the binary classification tasks of high/low Valence and high/low Arousal.
For the pre-training phase of the Automatic Piano Transcription model, we utilized the MAESTRO dataset ("MIDI and Audio Edited for Synchronous TRacks and Organization") \cite{onsets2}, encompassing a comprehensive collection of more than 200 hours of meticulously paired audio and MIDI recordings.

\begin{table}[]
\centering
\caption{Acoustic Encoder Settings.}
\label{setting}
\setlength{\tabcolsep}{2mm}{
\begin{tabular}{ccccc}
\toprule
\multirow{2}{*}{Layer} & \multirow{2}{*}{Channel} & \multirow{2}{*}{Kernel Size} & \multirow{2}{*}{Stride} & \multirow{2}{*}{Maxpooling} \\
                        &                          &                              &                         &                             \\ \midrule \noalign{\vskip 1mm}
conv1                   & 64                       & 3x3                          & 1                       & 2x2                         \\ \noalign{\vskip 1mm}
conv2                   & 128                      & 3x3                          & 1                       & 2x2                         \\ \noalign{\vskip 1mm}
conv3                   & 256                      & 3x3                          & 1                       & 2x2                         \\ \noalign{\vskip 1mm} \bottomrule
\end{tabular}
}
\end{table}


During the training process, the training data is divided into mini-batches with a batch size of 64. The Adam optimizer \cite{kingma2014adam} is employed, utilizing a learning rate of 0.0001. To implement all experiments, the PyTorch framework \cite{paszke2019pytorch} is utilized.

It is important to note that MIDI files from the EMOPIA dataset were not utilized in our experiments. As our proposed model exclusively takes audio files as input, our aim is to evaluate the overall performance of the complete model, including the refined AMT module.

\begin{table}
\centering
\caption{Comparison with symbolic-domain methods on EMOPIA. }
\label{comp1}
\setlength{\tabcolsep}{4mm}{
\begin{tabular}{cccc}
\toprule
\multirow{2}{*}{Method}                             & \multirow{2}{*}{4Q} & \multirow{2}{*}{A} & \multirow{2}{*}{V} \\
                                                    &                     &                    &                    \\ \midrule \noalign{\vskip 1mm}
LSTM-Attn \cite{emopia49}+MIDI-like \cite{emopia39} & .684                & .882               & .833               \\ \noalign{\vskip 1mm}
LSTM-Attn \cite{emopia49}+REMI \cite{emopia34}      & .615                & \textbf{.890}      & .746               \\ \noalign{\vskip 1mm}
symbolic-LR \cite{emopia}                           & .581                & .849               & .651               \\ \noalign{\vskip 1mm}
MIDIBERT \cite{midibert}                            & .634                & /                  & /                  \\ \noalign{\vskip 1mm}
MT-MIDIBERT \cite{symbolic,midibert}                & .676                & /                  & /                  \\ \noalign{\vskip 1mm}\cdashline{1-4}[1pt/1pt]\noalign{\vskip 1mm} \noalign{\vskip 1mm}
proposed model                                      & \textbf{.708}       & .874               & \textbf{.869}      \\ \noalign{\vskip 1mm} \bottomrule
\end{tabular}
}
\end{table}

\subsection{Comparative Studies}
We compared our proposed model with other existing methods on the same EMOPIA dataset.
To the best of our knowledge, there is no existing multi-domain piano emotion recognition research. So we compared our model with several uni-domain symbolic-domain models proposed in \cite{emopia} and \cite{symbolic}, including two models based on BLSTM and self-attention mechanism (LSTM-Attn for short) using MIDI-like and REMI symbolic representation, a linear regression model based on hand-crafted features, and two pre-trained Bert-like models. For a fair comparison, we directly used the original results announced in their works. In \cite{midibert,symbolic}, valence metrics and arousal metrics are not provided, therefore are not shown in the table.

Table \ref{comp1} shows the comparison between our method and the other five symbolic-domain methods. All the methods show high and similar performance on Arousal detection, which indicates that Arousal detection is a relatively simple task. Due to the strong sequence-modeling ability of our transformer-based symbolic domain model, our method shows the highest Valence detection performance and outperforms the LSTM-Attn+MIDI-like model by 3.6\%. On 4Q classification metrics, our model also achieves state-of-the-art performance and outperforms the LSTM-Attn+MIDI-like model by 2.4\%.  

We also compared our model with two existing acoustic-domain models, one uses linear regression on hand-crafted features and the other uses a ResNet-like network. Table \ref{comp2} shows the comparison between our method and the other two acoustic-domain methods. All acoustic-domain methods show strong performance on Arousal detection as well. 
This is in line with common sense, because Arousal is greatly affected by energy, velocity, and dynamics, and this information is evident in acoustic information. Though our method is slightly weaker on Arousal detection, it still outperforms the Short-chunk ResNet model by 3.1\% on the 4Q metrics.

\begin{table}
\centering
\caption{Comparison with acoustic-domain methods on EMOPIA.}
\label{comp2}
\setlength{\tabcolsep}{4mm}{
\begin{tabular}{cccc}
\toprule
\multirow{2}{*}{Method}                   & \multirow{2}{*}{4Q} & \multirow{2}{*}{A} & \multirow{2}{*}{V} \\
                                          &                     &                    &                    \\ \midrule \noalign{\vskip 1mm}
Audio-LR \cite{emopia}                    & .523                & \textbf{.919}      & .558               \\ \noalign{\vskip 1mm}
Short-chunk ResNet \cite{emopia50,resnet} & .677                & .887               & .704               \\ \noalign{\vskip 1mm}\cdashline{1-4}[1pt/1pt]\noalign{\vskip 1mm} \noalign{\vskip 1mm}
proposed model                            & \textbf{.708}       & .874               & \textbf{.869}      \\ \noalign{\vskip 1mm} \bottomrule
\end{tabular}
}
\end{table}

\subsection{Ablation Studies}
We designed and carried out a series of ablation studies to test the effect of our improvements. In the symbolic-only model and acoustic-only model, we use our symbolic branch and acoustic branch individually in order to test the effect of combining them. In the STFT-input model, we use an STFT spectrogram as input instead of the mixed acoustic feature. In the Single-loss model, we do not calculate the extra loss on the two branches and only calculate the Label loss.

The experimental results of the ablation studies are shown in Table \ref{ablation}. Compared to the two uni-domain models, our cross-domain fusion strategy costs performance loss on Arousal and Valence detection. However, our proposed model outperforms these two models by over 5\% on the overall 4Q accuracy metrics. This indicates that our model is able to make better decisions by considering both symbolic and acoustic information.

\begin{table}
  \centering
  \caption{Ablation studies trained and evaluated on the EMOPIA dataset.}
  \label{ablation}
  \setlength{\tabcolsep}{7mm}{
  \begin{tabular}{cccc}
  \toprule
  \multirow{2}{*}{Method} & \multirow{2}{*}{4Q} & \multirow{2}{*}{A} & \multirow{2}{*}{V} \\
                          &                     &                    &                    \\ \midrule \noalign{\vskip 1mm}
  Symbolic-only           & .651                & .843               & \textbf{.891}      \\ \noalign{\vskip 1mm}
  Acoustic-only           & .630                & \textbf{.902}      & .697               \\ \noalign{\vskip 1mm}
  STFT-input              & .689                & .804               & .871               \\ \noalign{\vskip 1mm}
  Single-loss             & .683                & .845               & .883               \\ \noalign{\vskip 1mm}
  proposed model          & \textbf{.708}       & .874               & .869               \\ \noalign{\vskip 1mm} \bottomrule
  \end{tabular}
  }
  \end{table}

The STFT-input model shows huge performance loss on Arousal metrics, which proves that using mixed acoustic features can improve Arousal detection performance. When using the STFT spectrogram as input, a deeper network is needed to extract the acoustic features. By using hand-crafted features, our method shows strong acoustic modeling ability with only three Conv layers. The single-loss model also shows over 2.5\% performance loss on both 4Q and Arousal metrics, which indicates that our strategy of calculating the extra loss function works.
     		

\section{Conclusion}
In this study, we introduce a novel multi-domain approach for piano emotion recognition, which can also be extended to other instruments with automatic transcription capabilities. Our proposed model leverages a pre-trained transcription model, enabling multi-domain analysis solely based on audio input. To the best of our knowledge, there is a lack of research specifically addressing piano emotion recognition. Our proposed model capitalizes on the complementary and redundant aspects between the acoustic and symbolic domains, leading to improved consistency in valence detection and arousal detection. Experimental results demonstrate that our proposed model surpasses the baseline approaches in terms of Valence classification and 4Q classification metrics. Moving forward, our future work will focus on designing enhanced symbolic representations for music, investigating superior cross-domain fusion strategies to enhance overall performance, and developing a universal framework for addressing the emotional aspects of transcribed musical instruments.

\section{Acknowledgement}
This paper is supported by the Key Research and Development Program of Guangdong Province under grant No.2021B0101400003. Corresponding author is Jianzong Wang from Ping An Technology (Shenzhen) Co., Ltd (jzwang@188.com).

\bibliographystyle{splncs04}
\bibliography{my,0-citation-self}

\end{document}